\documentclass[prl,twocolumn,superscriptaddress]{revtex4}
\usepackage{amsmath}
\usepackage[dvips]{graphicx}
\usepackage{bm}
\usepackage{epsfig}
\begin{document}

\title{Lasercooled RaF as a promising candidate to measure molecular parity violation}
\author{T.A.\ Isaev}
\email{isaev@fias.uni-frankfurt.de}
\affiliation{Clemens-Sch{\"o}pf Institute, 
             TU Darmstadt, Petersenstr. 22, 64287 Darmstadt, Germany}
\author{S.\ Hoekstra}
\email{s.hoekstra@rug.nl}
\affiliation{KVI, University of Groningen, Zernikelaan 25, 
             9747 AA Groningen, The Netherlands}       
\author{R.\ Berger}
\email{robert.berger@tu-darmstadt.de}
\affiliation{Clemens-Sch{\"o}pf Institute, 
             TU Darmstadt, Petersenstr. 22, 64287 Darmstadt, Germany}
%\graphicspath{./fig/}

\begin{abstract}

The parameter $W_\mathrm{a}$, which characterizes 
nuclear spin-dependent parity
violation effects within the effective molecular spin-rotational
Hamiltonian, was computed for the electronic ground state of radium
fluoride (RaF) and found to be one of the largest absolute values predicted
so far. These calculations were performed with the complex generalised
Hartree-Fock method within a two-component (quasi-relativistic)
zeroth-order regular approximation framework. Peculiarities of 
the molecular electronic structure of RaF lead to highly
diagonal Franck-Condon matrices between vibrational states of the
electronic ground and first excited states, which renders the molecule 
in principle suitable for direct laser cooling.
As a trapped gas of cold molecules offers a superior coherence time, 
RaF can be considered a promising candidate for high-precision spectroscopic 
experiments aimed at the search of molecular parity-violation effects.
\end{abstract}

\maketitle

%=============================================================================
Diatomic radicals containing heavy nuclei are, together with closed shell
chiral molecules with high $Z$ nuclei \cite{laubender:2003,Nahrwold:09},
especially suitable for the search of %(space) 
parity violation effects %(P-odd effects) 
which depend on the nuclear spin (see review
\cite{Kozlov:95}). Up to the present time, the only measured nuclear-spin dependent parity violating (NSDPV) quantity is the anapole moment \cite{zeldovich:1957,zeldovich:1958} of the $^{133}$Cs nucleus \cite{Wood:97}. Nevertheless, it was concluded that the result of this
experiment is in contradiction with previously obtained restrictions on
NSDPV constants (see review \cite{Ginges:04}), which calls for further
attempts to determine NSDPV effects. Currently, an experiment on measurement of the nuclear 
anapole moment of barium in the barium fluoride
(BaF) molecule using a molecular beam is under preparation at Yale University
\cite{DeMille:08}. 
Here we propose the use of trapped radium fluoride (RaF) molecules as a promising candidate to probe parity violation, which has never been detected in molecules. 

In recent years a number of techniques has been developed to prepare, control and cool molecules~\cite{Carr:2009}. A level is now being reached where these techniques can have a significant impact on precision experiments. Stark deceleration, well suited to slow down and trap diatomic polar molecules, has been developed in the last decade~\cite{Bethlem:1999, Bethlem:2000, Meerakker:2008} and the possibility of laser cooling of molecules has only very recently received more attention~\cite{DiRosa:04, Stuhl:08, Shuman:09}. If such cooling methods can be successfully applied to sensitive molecules, a new generation of precision measurements can be done.

As we show in this paper, the RaF molecule appears very well suited for
laser cooling methods, because of 1) the highly diagonal Frank-Condon matrix
between the electronic ground state and the lowest electronically excited
state as well as 2) the facts i) that the corresponding electronic transition
frequency is in the visible region with a reasonable lifetime and ii) that
isotopes with nuclear spin quantum number $I=1/2$ are available. Combined
with its large $W_\mathrm{a}$, this makes the RaF molecular a prime
candidate for a precision experiment probing molecular parity violation.

An experiment with radioactive molecules requires a specialized facility. At the KVI of the University of Groningen the TRI$\mu$P facility on the production of radioactive isotopes was commissioned in the year 2007, with the perspective of using these radioactive isotopes in precision experiments to test fundamental symmetries~\cite{Traykov:08}. As the accuracy in a spectroscopic measurement is ultimately limited by the coherence time, there is a clear potential for precision experiments using trapped species. A number of precision experiments using trapped species has been set up, such as Ra$^+$ to probe parity violation~\cite{Wansbeek:2008} and Ra atoms for electron electric dipole moment (eEDM) detection~\cite{De:09}. We will discuss the production, deceleration and trapping of RaF molecules furtheron in this paper.

%=============================================================================
\paragraph{Calculation of the $W_\mathrm{a}$ parameter}
%=============================================================================
Previously \cite{Isaev:10} we utilised the quasi-relativistic two-component (2c)
zero order regular approximation (ZORA) for calculating the NSDPV parameter
$W_\mathrm{a}$ of the effective molecular spin-rotational Hamiltonian in a
number of heavy-atom diatomic molecules. The corresponding P-odd
contribution to this Hamiltonian reads as $W_\mathrm{a} k_A
[\vec{\lambda}{\times}\vec{S}_\mathrm{eff}] \cdot \vec{I}$ with the
effective electron spin operator $\vec{S}_\mathrm{eff}$, the nuclear spin
operator $\vec{I}$, the vector $\vec{\lambda}$ which points from the heavy
nucleus to the light one and the parameter $k_{A}$ of the heavy nucleus
\cite{Ginges:04} that includes the effect of the nuclear anapole moment and
the NSD $Z^{0}$ exchange contribution. One may additionally attempt to
absorb effects from the nuclear-spin independent $Z^{0}$ exchange
contribution in $k_{A}$, but as this contribution depends via the hyperfine
coupling tensor also on the specific electronic structure, it may appear
more canonical to include this contribution in the electronic structure
parameter $W_\mathrm{a}$. For the time being, however, we have omitted this
(supposedly small) additional term. 

The results obtained with this approach
are at the level of paired generalised Hartree-Fock (paired GHF, for the
classification used see review \cite{Stuber:03}) for
\emph{off-diagonal} matrix elements (MEs) between the degenerate electronic
states while \emph{diagonal} MEs correspond to complex GHF, thereby
including spin-polarisation effects. The advantage of our approach is that
it can be straight-forwardly extended to polyatomic molecules. In the case
of diatomic molecules with $^2\Sigma$ ground state, we can obtain reliable
estimates of the NSDPV effect together with the hyperfine coupling tensor
$\mathbf{A}$ using ZORA. In our calculations we used a modified version
\cite{berger:2005,berger:2005a,Isaev:10} of the {\sc TURBOMOLE} code
\cite{ahlrichs:1989,haser:1989} with an even-tempered atomic basis set on
Ra, a basis set of triple-zeta quality from the {TURBOMOLE} basis set
library augmented with diffuse p- and d-functions on F atom 
(see supplementary material at [URL will be inserted by AIP] for
basis sets and other parameters of the calculations).
A finite spherical Gaussian nucleus model is used and a ZORA model
potential as proposed by van W{\"u}llen \cite{wullen:1998} with additional
damping \cite{liu:2002} was applied.  
The equilibrium distance $R_\mathrm{e}$ of the electronic ground state was
obtained from complex GHF-ZORA calculations and found to be in good
agreement with $R_\mathrm{e}$ determined in later four-component
(relativistic) Fock space coupled cluster singles and doubles (FS-CCSD)
calculations. As one can see from the Table \ref{Wa} the changes for all
parameters are within 6~\% when the $R_\mathrm{e}$ value is taken either
from FS-CCSD or GHF-ZORA calculations. The computed absolute value of the
$W_\mathrm{a}$ parameter is in the kHz range and thus one of the largest
predicted for diatomic molecules so far.
As Ra nuclei have an even number of protons, measurements of $W_\mathrm{a}$ in RaF would complement ideally previous ${}^{133}$Cs experiments on the nuclear anapole moment of nuclei
with an odd number of protons. The hyperfine coupling tensor calculated 
herein for RaF with accounting for spin-polarisation effects is highly 
isotropic and can be used, once measured, for estimating the quality of 
our prediction for $W_\mathrm{a}$.
%------------- Table 1 ----------------------------
\begin{table}[h]
\caption
 {Molecular parameters for the 2c-GHF-ZORA description of 
 the electronic ground state $^2\Sigma_{1/2}$ of $^{223}$RaF and
 $^{225}$RaF.
 The results are given for both $R_\mathrm{e}$ taken from the GHF-ZORA ($4.40~a_0$) 
 and FS-CCSD ($4.24~a_0$) calculations. The parallel and perpendicular components of the 
 hyperfine tensor are given in GHz, $W_\mathrm{a}$ in kHz (with estimated dependence 
 on the Ra mass number being on the order of a few Hz).
 \label{Wa}
}
%\vspace{0.2cm}
\begin{tabular}{l|cc|cc|c}
 \multicolumn{6}{c}{}\\
 \multicolumn{1}{c}{} &\multicolumn{2}{c}{$^{223}$RaF}&\multicolumn{2}{c}{$^{225}$RaF} & \\
$R_\mathrm{e}/a_0$ & $A_\parallel$/GHz & $A_\perp$/GHz & $A_\parallel$/GHz & $A_\perp$/GHz 
& $W_\mathrm{a}$/kHz   \\ 
\hline
  &  &  & &  & \\
 4.40  & ~1.86~~  & ~1.82~~  & ~$-15.1$~~  & ~$-14.8$~~  & ~1.36~~  \\
  &  &  & &  & \\
  &  &  & &  & \\
 4.24     & ~1.90~~ &  ~1.86~~  & ~$-15.4$~~ &~$-15.1$~~  &  ~1.30~~ \\
  &  &  & &  & \\
 
\end{tabular}
\\
\end{table}
%--------------------------------------------------

%--------------------------------------------------

Besides the high $W_\mathrm{a}$ value there is also another peculiarity which 
makes RaF very attractive for experimental NSDPV search, namely that this 
molecule appears well suited for being directly cooled with lasers.
%end 
%=============================================================================
\paragraph{Franck-Condon factors and direct laser cooling of molecules}
%=============================================================================
One of the major problems in direct laser cooling of molecules is the
spontaneous radiative decay into a manifold of rovibrational states not
covered by the cooling cycle. 
Transitions between rotational states are driven by selection rules for vector operators and can be
reduced to the two-level case \cite{Stuhl:08}, but depopulation of
vibrational levels during cooling cycles is the obstacle. In
Ref.~\cite{DiRosa:04} several molecules were proposed which have highly
diagonal overlap matrices between vibrational eigenfunctions of the ground
and the first excited electronic states (in terms of the molecular
spectroscopy --- highly diagonal Franck-Condon matrices).  For $10^4$ to
$10^5$ cooling cycles, the probability of the molecule to stay inside the
cooling loop has to be greater than 0.9999 with the \emph{fourth digit being
significant}. In practice this means that one has to know FC factors with
an accuracy of 0.01\%, which is extremely challenging to achieve in modern
calculations (although FC factors can in principle be measured with high
accuracy). Thus \emph{predictions} of FC factors with the accuracy required
by experimental needs appear practically impossible. 

Nevertheless, one can try to narrow down the number of candidate systems
for direct cooling with laser based on peculiarities of the molecular
electronic structure which renders certain molecules promising. We have
analysed the electronic structure of diatomic molecules with one electron
over closed shells. Such molecules have large magnetic moments
(in comparison with the closed shell systems)
which makes them suitable for magnetic trapping. More detailed considerations
shall be made in a later article, here we would just like to give the short
summary. Highly diagonal FC matrices should be expected when the unpaired electron 
is making transitions between orbitals that do not contribute to the bonding.
Such one-electron wavefunctions can be of the following type:
%\begin{enumerate} 
%
%\item 
{\bf 1)} ``Electron in lone orbital''. This situation is well-known in
theoretical chemistry, when, \emph{e.g.} as a result of sp$^n$ hybridisation, 
an unpaired electron is directed out of the bonding region.
This takes place \emph{e.g.} in the electronic ground state of 
MF and MH, where M
belongs to the group of alkaline earth metals (Mg to Ra). 
Two molecules from this series were already used to indicate the perspectives of
direct molecular cooling, namely CaH in Ref.~\cite{DiRosa:04} and SrF in
Ref.~\cite{Stuhl:08}.
%
%\item 
{\bf 2)} ``Atom-like electron in molecule''. In this case, the outer electron
is, due to some reasons (symmetry as a rule), located in a
highly-symmetrical state, which has practically atomic nature. Such a
situation happens for example in HI$^+$ \cite{Isaev:05a}, where the
unpaired electron resides in an orbital which is mainly of atomic $p_{3/2}$
character centered on iodine.  
In isoelectronic neutral
molecules (such as TeH) the situation is expected to be the same.
%
%\item 
{\bf 3)} In heavy-atom compounds, the valence electron orbital is often quite
diffuse, thus the maximum density of the unpaired electron is located
further away from the molecular core than in lighter homologues.
Thus among all possible molecules, those with heavier nuclei appear
promising.
%
%\end{enumerate}

The heavy-atom molecules RaF, HgH, HgF etc. which should
belong to the first class (``Electron in lone orbital'') and HI$^+$, PoH, 
TeH etc. which should belong to the second class (``Atom-like electron in 
molecule'') are expected to have highly diagonal FC matrices and, thus,
appear, in principle, highly suitable candidates for direct cooling of 
molecules. 
Unfortunately, the molecules belonging to the second class are not
particularly suitable for NSDPV experiments as the unpaired valence
electron is in an orbital with high total
%tim
angular
%end
momentum and does not have
noticeable spin density in the vicinity of heavy nucleus. As to the first
class, the molecules HgH and HgF have (as RaF) a $^{2}\Sigma$ ground state
and, according to our estimates, even higher $W_\mathrm{a}$ values (for HgH
it is 3.3 kHz,
%tim
although we would like to emphasize again that spin-polarisation and other electron 
correlation effects are not accounted in our $W_\mathrm{a}$ estimates, 
while these effects can well be important).
%end
The excitation energy of the first
excited state in these molecules, however, is in violet or ultraviolet region 
(for HgH 407 nm, for HgF 256 nm) \cite{Huber:79}. Especially at 256 nm it is difficult to produce sufficiently intense laser light.
In the series MF, instead, where M belongs to the group of alkaline earth metals, the excitation energies from the ground to first excited electronic states are in the visible region, so one
can expect a similar situation in RaF. This wavelength region is directly accessible using diode lasers.

To substantiate the considerations above, we performed exemplary
four-component Fock space (FS) CCSD calculations of the six lowest-lying
states of RaF using the {\sc DIRAC08} package \cite{DIRAC:08}. 
The details of the FS-CCSD approach are given in the review \cite{Visscher:01}.
Electrons of the $6$s,$6$p,$7$s shells of Ra and $2$s,$2$p of F are correlated, 
the basis sets and other computational parameters can be located in the
supplementary material.

\begin{figure}[h]
  \centering
  \includegraphics[width=\linewidth]{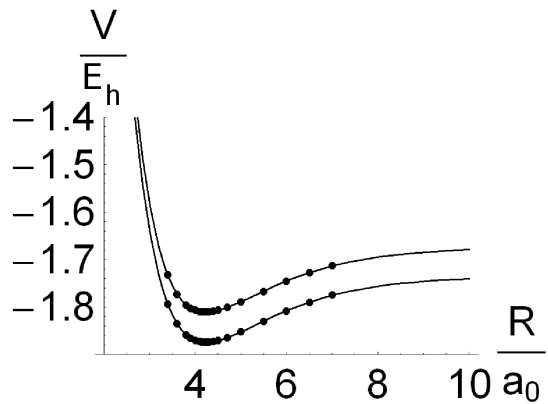}
  \caption{Morse potential fitting of the calculated 
           potential energy points for the two 
           energetically lowest-lying 
           electronic states of RaF -- the ground electronic state is 
           $^2\Sigma_{1/2}$ and first electronically excited state is
           $^2\Pi_{1/2}$.}
  \label{plot}
\end{figure}

The two energetically lowest lying electronic states are identified 
as $^2\Sigma_{1/2}$ (ground state) and $^2\Pi_{1/2}$ (first excited state). 
Then the points calculated in the interval $3.5~a_0$ to $7.0~a_0$ were 
fitted by a Morse potential (the points and fitting curves 
are given in Figure \ref{plot}) and spectroscopic parameters together with the
corresponding FC factors were determined. The FC factors were calculated
by the {\sc Mathematica} script for FC factors calculation between 
vibrational Morse oscillator states taken from Ref.~\cite{Lopez:02}. The results obtained are 
presented in Table~\ref{raf}. 

The laser cooling transition ($^2\Pi_{1/2}$($v^\prime$=0) $\leftarrow$ $^2\Sigma_{1/2}$($v$=0))  is found to lie around 710 nm, and an upper state lifetime on the order of $\sim 25$ ns is expected. With these parameters the capture range for optical molasses is $\sim 5$ m/s, and the doppler limit temperature is $\sim 150 \mu$K. If a precooled sample of molecules can be created, as is described in the next section, only a few thousand photons would have to be scattered per molecule to reach this temperature. At these temperatures only a very small volume in the trap is occupied, opening the possibility to perform precision measurements using the trapped molecules.

The sum of the FC factors for the ground vibrational state of 
the $^2\Pi_{1/2}$ state and three vibrational states of $^2\Sigma_{1/2}$ state 
is equal to $0.9999$ with four digits after a point being stable 
in respect to variations of the fitting parameters of Morse potential
(see supplementary materials for details).
We would like to emphasize here, that high \emph{digital} precision in calculations 
of FC factors 
does not mean high accuracy of the spectroscopic constants calculations. 
Reliable estimate of the sum of FC factors which can be used for 
practical purposes is expected to have two digits after a decimal point.
Nevertheless \emph{both} numerical results and arguments from electronic structure theory provide, in our 
opinion, good reasons for considering RaF for measurements of NSDPV effect and for direct laser cooling.
%end
%------------- Table 1 ----------------------------
\begin{table}[h]
\caption
 {Estimated molecular spectroscopic parameters for the
  electronic ground state $^2\Sigma_{1/2}$ and the first electronically
  excited $^2\Pi_{1/2}$ state from FS-CCSD calculations of RaF. The dissociation 
  energy $\tilde{D}$ was obtained as the difference between total energies 
  of the states at the equilibrium distance $R_e$ and at $9.0~a_0$ without 
  attempting to correct for basis set superposition errors. 
  Fitting parameters of Morse potential with
  the corresponding standard errors can be found in supplementary materials. 
  Parameter $\tilde{\omega}_\mathrm{e}$ is the standard harmonic vibrational wavenumber.
  We would like to emphasize that the accuracy of the determination of the parameters is
  lower (especially for $\tilde{D}$ and $\tilde{T}_\mathrm{e}$) than 
  the number of digits reported. Nevertheless we report the 
  results with high digital precision to provide reaper points for 
  future calculations.} 
 \label{raf}
%\vspace{0.2cm}
\begin{tabular}{lcccc}
 \vspace{3mm} & $R_\mathrm{e}/a_0$ & $\tilde{\omega}_\mathrm{e}/\mathrm{cm}^{-1}$  &
$\tilde{D}_\mathrm{e}/\mathrm{cm}^{-1}$     & $\tilde{T}_\mathrm{e}/\mathrm{cm}^{-1}$ \\

%& & & & \\
\hline
 \\
$^2\Sigma_{1/2}$    &   4.24   &    428           &    32104             &  \\
                    &             &                  &           & \\
$^2\Pi_{1/2}$       &  4.24    &    432            & 31302 &  14012  \\
     &    &   &  &\\ \hline
\end{tabular}
\\
\end{table}
%--------------------------------------------------------------
%============================================================================
\paragraph{Proposed measurement strategy and sensitivity to NSDPV signal}
%============================================================================
%[main idea]
The $W_\mathrm{a}$ parameter can be obtained in a diatomic molecule by a measurement of the parity-violating coupling between hyperfine-rotational levels of opposite parity. An experimental procedure to do so in a molecular beam has been proposed in Ref~\cite{DeMille:08},
where tuning of the close-lying spin-rotational levels of opposite parity 
to near-degeneracy due to the Zeeman effect is used to increase the P-odd asymmetry 
in the measured signal.
The accuracy in the determination of $W_\mathrm{a}$ that can be reached in such an experiment is proportional to the count rate of the molecular signal ($dN/dt$), the interaction (coherence) time ($T$) and the total measurement time ($\tau$) as $\delta W_\mathrm{a} \sim 1/(T\sqrt{\tau dN/dt})$.

For an interaction region of 5 cm and a molecular beam velocity of 500 m/s, the interaction time is on the order of 100 $\mu$s. As it has been demonstrated that cold molecules can be trapped for at least a number of seconds~\cite{Hoekstra:2007, Vanhaecke:2007a}, the potential gain in accuracy is 4 orders of magnitude. It is also clear that part of this gain is lost as the number of molecules will probably be low; 
so production and delivery of the RaF molecules to the working volume is crucial for 
the proposed measurements. On the other hand the high Franck-Condon overlap allows 
for the detection of multiple photons even from a single trapped molecule.

The two radium isotopes that are most suitable are $^{213}$Ra and $^{225}$Ra, with lifetimes of 164 s and 14.9 days, respectively. Both isotopes have a nuclear spin $I=1/2$, which reduces the number of lasers required for lasercooling. The $^{213}$Ra isotope has been successfully produced and used in experiments at the KVI~\cite{Versolato:2010a}, whereas $^{225}$Ra is best obtained from a $^{229}$Th source. 
The radium ions from the cyclotron can be implanted in a solid AlF$_3$ target. RaF molecules can be produced from such a target by laser ablation. Similar molecules have been obtained in the rovibriational groundstate by the use of a cold buffer gas or supersonic expansion~\cite{Tarbutt:2009,Tarbutt:2002}. These precooling methods greatly reduce the amount of photons that have to be scattered in order to reach the doppler temperature, which is a large advantage if the cooling cycle is not completely closed.

A supersonic beam can be combined with a Stark decelerator to bring the molecules to rest in the lab frame. A fraction of about $10^{-2}$ of the molecules in the supersonic expansion can be trapped. Heavy diatomic molecules such as RaF have not been decelerated to standstill yet; but a recent demonstration of an improvement to the Stark-deceleration technique~\cite{Osterwalder:2010} indicates how this could be done. The efficiency of capturing the ablated molecules in the supersonic beam is however only about $10^{-8}$ 
\cite{Willmott:2000,Tarbutt:2002}, which is prohibitively low for RaF.

If the ablation is performed in a buffer gas cell, the cold molecules can be captured by a magnetic trap~\cite{Weinstein:1998}. The buffer gas has to be quickly removed~\cite{Harris:2004a}, and subsequently lasercooling could be used to bring the trapped cloud to the lower temperatures that are required for a precision measurement. A promising number of $10^{12}$ YbF molecules, produced in a single laser ablation shot, has been reported~\cite{Tarbutt:2009} but it remains to be seen how many of these can eventually be trapped and used for a precision measurement. 

%============================================================================
%\paragraph{Conclusion and outlook}
%============================================================================
To conclude, we have investigated the suitability of the radium fluoride (RaF) molecule to probe parity violation, which has so far not been detected in molecules.
Explicit {\it ab initio} calculations of the parity violating nuclear spin dependent
parameter $W_\mathrm{a}$ and Franck-Condon factors for transitions between
the electronic ground state and the lowest doublet excited state indicate
that this open-shell molecule is a promising candidate for
direct laser cooling and (subsequent) measurement of the nuclear anapole
moment. 
We have also discussed possible approaches to creation, trapping and cooling of 
RaF molecules.
In our opinion RaF is a valuable addition to the variety of molecules that 
may be used to probe parity violation, providing complimentary insight 
on the NSDPV effects.

%============================================================================
\paragraph*{Acknowledgements}
%============================================================================
We are indebted to Sophie Nahrwold, Mikhail Kozlov, Klaus Jungmann and Rob Timmermans 
for discussions.
Financial support by the Volkswagen Foundation and computer time provided
by the Center for Scientific Computing (CSC) Frankfurt is gratefully
acknowledged.
%------------- Figure ------------------------

\end{document}